\documentclass[3p,times,fleqn]{elsarticle}

\usepackage[ruled,vlined]{algorithm2e}
\usepackage[numbers]{natbib}
\usepackage{amsmath,amsfonts}
\usepackage{xcolor}

\usepackage{siunitx}

\begin{document}

\title{An eikonal model with re-excitability for fast simulations in cardiac electrophysiology}


\author[1]{Lia Gander}
\ead{lia.gander@usi.ch}
\address[1]{Euler Institute, Università della Svizzera italiana,
            via G.~Buffi 13, CH-6900 Lugano, Switzerland}

\author[2,7]{Rolf Krause}
\ead{rolf.krause@fernuni.ch}
\address[2]{FernUni,
            Schinerstrasse 18, CH-3900 Brig, Switzerland}
\address[7]{AMCS, CSE Department, KAUST, Thuwal-Jeddah, Saudi Arabia}

\author[3,4,5]{Francisco Sahli~Costabal}
\ead{fsc@ing.puc.cl}
\address[3]{Department of Mechanical and Metallurgical Engineering, School of Engineering, Pontificia Universidad Cat\'olica de Chile, Santiago, Chile}

\address[4]{Institute for Biological and Medical Engineering, Schools of Engineering, Medicine and Biological Sciences, Pontificia Universidad Cat\'olica de Chile, Santiago, Chile}

\address[5]{Millennium Institute for Intelligent Healthcare Engineering, iHEALTH}

\author[1,6]{Simone Pezzuto\corref{cor1}}
\ead{simone.pezzuto@unitn.it}
\address[6]{Laboratory of Mathematics for Biology and Medicine,
            Universit\`a di Trento, via Sommarive 14, I-38123 Trento, Italy}
\cortext[cor1]{Corresponding author}

\begin{abstract}
Precision cardiology based on cardiac digital twins requires accurate simulations of cardiac arrhythmias.
However, detailed models, such as the monodomain model, are computationally costly and have limited applicability in practice. Thus, it desirable to have fast models that can still represent the main physiological features presented during cardiac arrhythmias. 
The eikonal model is an approximation of the monodomain model that is widely used to describe the arrival times of the electrical wave.
However, the standard eikonal model does not generalize to the complex re-entrant dynamics that characterize the cardiac arrhythmias.
In this work, we propose an eikonal model that includes the tissue re-excitability, which allows to describe re-entries.
The re-excitability properties are inferred from the monodomain model.
Our eikonal model also handles the tissue anisotropy and heterogeneity.
We compare the eikonal model to the monodomain model in various numerical experiments in the atria and the ventricles.
The eikonal model is qualitatively accurate in the simulation of re-entries and can be potentially ran in real-time, opening the door to its clinical applicability.

\end{abstract}

\begin{keyword}
eikonal model \sep
monodomain model \sep
fast marching method \sep
cardiac arrhythmias \sep
atrial fibrillation
\end{keyword}

\maketitle

\section{Introduction}


Cardiac electrophysiology models are widely used for \textit{in-silico} patient-specific clinical investigations, e.g., the understanding and the treatment of arrhythmias, such as atrial fibrillation (AF) \cite{mcdowell2015virtual,boyle2019computationally,gharaviri2020epicardial,gharaviri2021synergistic} or ventricular tachycardia (VT) \cite{ashikaga2013feasibility,trayanova2017imaging,deng2019characterizing,gionti2022role}.
The most commonly used model is the monodomain model for the transmembrane potential. The underlying monodomain system consists of a parabolic reaction-diffusion equation coupled with a ionic ordinary differential equation system.
Accurate high fidelity computational models, such as the monodomain model, have high computational cost, which often hinders their applicability to personalized clinical applications, where the results need to be delivered within tight time constraints.

There are alternatives that are less accurate but faster to speed up the computations. These models can be approximate models based on simplified physics, such as the reaction-eikonal model \cite{neic2017efficient,nagel2023comparison} or the eikonal model \cite{wallman2012comparative,nagel2023comparison}, reduced-order models \cite{fresca2020deep,cicci2022projection}, or simply models based on a coarser discretization of the computational domain in the numerical solution of the high fidelity equations \cite{pagani2021computational,gander2022fast}.
In this work we focus on the eikonal model for the activation times. The underlying eikonal equation is only space dependent, has no diffusion term and does not describe the ionic properties. Therefore the numerical solution of the eikonal equation is cheaper than the numerical solution of the monodomain system.
Our goal is to develop an eikonal model for cardiac arrhythmias and assess its quality by comparing it to the monodomain model.

Cardiac arrhythmias are characterized by a self-sustained re-entrant electrical activity. The re-entries imply several consecutive activations of the tissue. The eikonal model describes the activation times of a single activation. Therefore, to model cardiac arrhythmias, we need to adapt the eikonal model to account for the tissue re-excitability and to allow for multiple activations.
The eikonal equation is typically solved by algorithms that iteratively pass through the nodes of a computational mesh and update the corresponding activation times until acceptance. These iterative algorithms are appropriate to include the necessary adaptations of the eikonal model for our goal. In particular, to include the re-excitability, a suitable iterative algorithm has to be single-pass and has to guarantee the monotone acceptance of the nodes with respect to their activation times. The fast marching method (FMM) \cite{kimmel1998computing,sethian2000fast} satisfies these requirements and is, therefore, our method of choice.

Our adaptation of the FMM defines a global time variable and a time-stepping that allow to alternate between activation times updates and updates of the restitution properties, which describe the tissue re-excitability.
Our restitution properties are computed from the monodomain model.
Moreover, our FMM approach handles the tissue anisotropy with mesh adaptation. Specifically, we use anisotropic meshes to guarantee the convergence of the FMM to the viscosity solution of the eikonal equation \cite{mirebeau2014anisotropic}. 
Additionally, the FMM also handles the tissue heterogeneity, e.g.~due to the structural remodeling.

In the literature, fast simulations of re-entries have often been carried out by combining the monodomain and the eikonal models.
In \cite{jacquemet2010eikonal} and \cite{jacquemet2012eikonal}, the monodomain model is combined with the eikonal-diffusion model to generate re-entries along prescribed pathways.
In \cite{corrado2018conduction}, a reaction-eikonal model is used to simulate a train of stimuli.
The same reaction-eikonal model is used in \cite{gassa2021spiral} to simulate spiral waves.
In \cite{barrios2022diffusion}, the monodomain and the eikonal models are alternated to generate a re-entry on a ring.
Recently, cellular automata have also been used to perform fast simulations of spiral waves \cite{romitti2023cellular} and VT \cite{serra2022automata}.

Few works use a pure eikonal approach to carry out fast simulations with re-excitability evaluations.
In \cite{corrado2018conduction}, an adapted Dijkstra's method is used to simulate a train of stimuli.
The FMM is adapted to simulate a train of stimuli in \cite{sermesant2007anisotropic}, to initiate VT in \cite{pernod2011multi} and to generate atrial flutter (AFlut) along precomputed re-entry pathways in \cite{loewe2019patient}. 
In this work, we use the FMM algorithm with re-excitability to simulate various re-entries and we propose a novel formal treatment of the anisotropy.

The rest of the manuscript is organized as follows. In Section~\ref{sec: model} we introduce the monodomain and the eikonal models, we compute the restitution curves of the monodomain model, and we present the FMM and its adaptation to include the re-excitability. In Section~\ref{sec: experiments} we present some numerical experiments, in particular the simulations of ventricular tachycardia and atrial flutter. A discussion follows in Section~\ref{sec: discussion}.

\section{Methods}
\label{sec: model}

\newcommand{\vx}{\boldsymbol{x}}
\newcommand{\vw}{\boldsymbol{w}}
\newcommand{\vc}{\boldsymbol{c}}
\newcommand{\vn}{\boldsymbol{n}}
\newcommand{\Dm}{\mathbf{D}_\mathrm{m}}
\newcommand{\Cm}{C_\mathrm{m}}
\newcommand{\Iion}{I_\mathrm{ion}}
\newcommand{\Iapp}{I_\mathrm{app}}

\subsection{The monodomain equation}

We start by introducing our reference model for cardiac electrophysiology, namely the monodomain equation~\cite{colli2014mathematical}. Consider a domain $\Omega\subset\mathbb{R}^3$, representing the active myocardium, and a time interval $[0,T]$.
The monodomain equation for the transmembrane potential $v\colon\Omega\times[0,T)\rightarrow\mathbb{R}$ reads as follows
\begin{equation}
\left\{ \begin{aligned}
& \nabla\cdot \bigl(\Dm \nabla v \bigr) = \beta\left( \Cm \frac{\partial v}{\partial t} + \Iion (v,\vw) - \Iapp \right)
&&\text{for $\vx \in \Omega$, $0 < t < T$}, \\
& \frac{\partial \vw}{\partial t} = \boldsymbol{G}(v,\vw),
&&\text{for $\vx \in \Omega$, $0 < t < T$}, \\
& \Dm\nabla v\cdot \vn = 0, 
&&\text{for $\vx \in \partial\Omega$, $0<t<T$}, \\
&   v(\vx,0)=  v_0(\vx), &&\vx\in\Omega, \\
& \vw(\vx,0)=\vw_0(\vx), &&\vx\in\Omega,
\end{aligned}\right.
\label{eq:monodomain}
\end{equation}
where $\Dm\colon\Omega\rightarrow\mathbb{R}^{3\times 3}$ is the monodomain conductivity tensor, symmetric positive-definite, $\Cm=\SI{1}{\micro\farad\per\square\cm}$ is the membrane capacitance, $\Iapp(\vx,t)$ is an applied transmembrane current per unit area, and $\beta=\SI{800}{\per\cm}$ the surface-to-volume ratio~\cite{gharaviri2020epicardial}. The nonlinear term $\Iion(u,\vw)$ is the ionic transmembrane current per unit area, and it depends on the type of tissue under consideration. The ionic current is determined by the transmembrane current $v$ and a possibly large set of auxiliary variables $\vw(\vx,t)$, which describe the gating of ion channels, the intracellular calcium dynamic, and other ionic concentration variables.
We define the monodomain conductivity tensor as transversely isotropic as follows:
\begin{equation*}
\Dm(\vx)=\sigma_\mathrm{m}^t(\boldsymbol{x})\boldsymbol{I}+\big(\sigma_\mathrm{m}^l(\boldsymbol{x})-\sigma_\mathrm{m}^t(\boldsymbol{x})\big)\boldsymbol{f}_l(\boldsymbol{x})\boldsymbol{f}_l(\boldsymbol{x})^\intercal,
\end{equation*}
where $\sigma_\mathrm{m}^{l,t}\colon\Omega\rightarrow\mathbb{R}$ are respectively the longitudinal and transversal conductivities defined as:
\begin{equation*}
\sigma_\mathrm{m}^t = \frac{\sigma_\mathrm{m}^{t,i}\sigma_\mathrm{m}^{t,e}}{\sigma_\mathrm{m}^{t,i}+\sigma_\mathrm{m}^{t,e}},\qquad
\sigma_\mathrm{m}^l = \frac{\sigma_\mathrm{m}^{l,i}\sigma_\mathrm{m}^{l,e}}{\sigma_\mathrm{m}^{l,i}+\sigma_\mathrm{m}^{l,e}},
\end{equation*}
and $\boldsymbol{f}_l\colon\Omega\rightarrow\mathbb{R}^3$ is the fiber orientation. The superscript $(\cdot)^{i,e}$ refers to the intracellular and extracellular conductivities respectively.

\subsection{Restitution curves}
\label{sub: restitution}

Cardiac arrhythmias are characterized by a re-entrant electrical activity that implies several activations.
These activations are allowed by the re-excitability of the tissue.
In this work, we include the restitution properties describing the re-excitability in the eikonal model, computed from the monodomain model.
In particular, we compute the restitution curves that determine the action potential duration (APD) and the conduction velocity (CV) from the diastolic interval (DI).

The procedure to compute the restitution curves consists of a prepacing at a basic cycle length (BCL) and a pacing protocol to determine the DI, the APD and the CV associated to various cycle lengths (CLs) \cite{xie2002electrical, cherry2008dynamics, wilhelms2013benchmarking}. Specifically, the protocol consists of a sequence of stimuli at BCL, followed by one stimulus at the considered CL, which is progressively decreased.
We compute the APD restitution curve in 0D with time step $\Delta t=\SI{0.02}{\ms}$. The APD is defined as the time interval during which the transmembrane potential $v$ is above the \SI{-62}{\milli\volt} threshold.
We compute the CV restitution curve with the same time step in a 1D tissue strand of length \SI{4}{\cm} and discretized with spatial resolution $h=\SI{0.05}{\milli\meter}$. The stimuli have size \SI{0.1}{\cm} and are delivered from one end of the strand. The CV is computed from the activation times at the center of the two halves of the domain. The activation time is defined as the time when $v$ reaches \SI{-62}{\milli\volt}.

In this work we consider three ionic models for three different tissue types.
For the atrial tissue, we consider the Courtemanche-Nattel-Ramirez ionic model of the human atria \cite{courtemanche1998ionic} adapted to describe the electric remodeling due to AF~\cite{gharaviri2020epicardial} and to reduce numerical instabilities~\cite{potse2019inducibility,Potse2019AF}.
For the ventricular tissue we consider two versions of the Mitchell-Schaeffer ionic model \cite{mitchell2003two}, one for the healthy tissue and one for the border zone surrounding the non-conductive scars that characterize the structurally remodeled tissue~\cite{pernod2011multi}.

\begin{figure}[t]
\centering
\includegraphics[width=\textwidth]{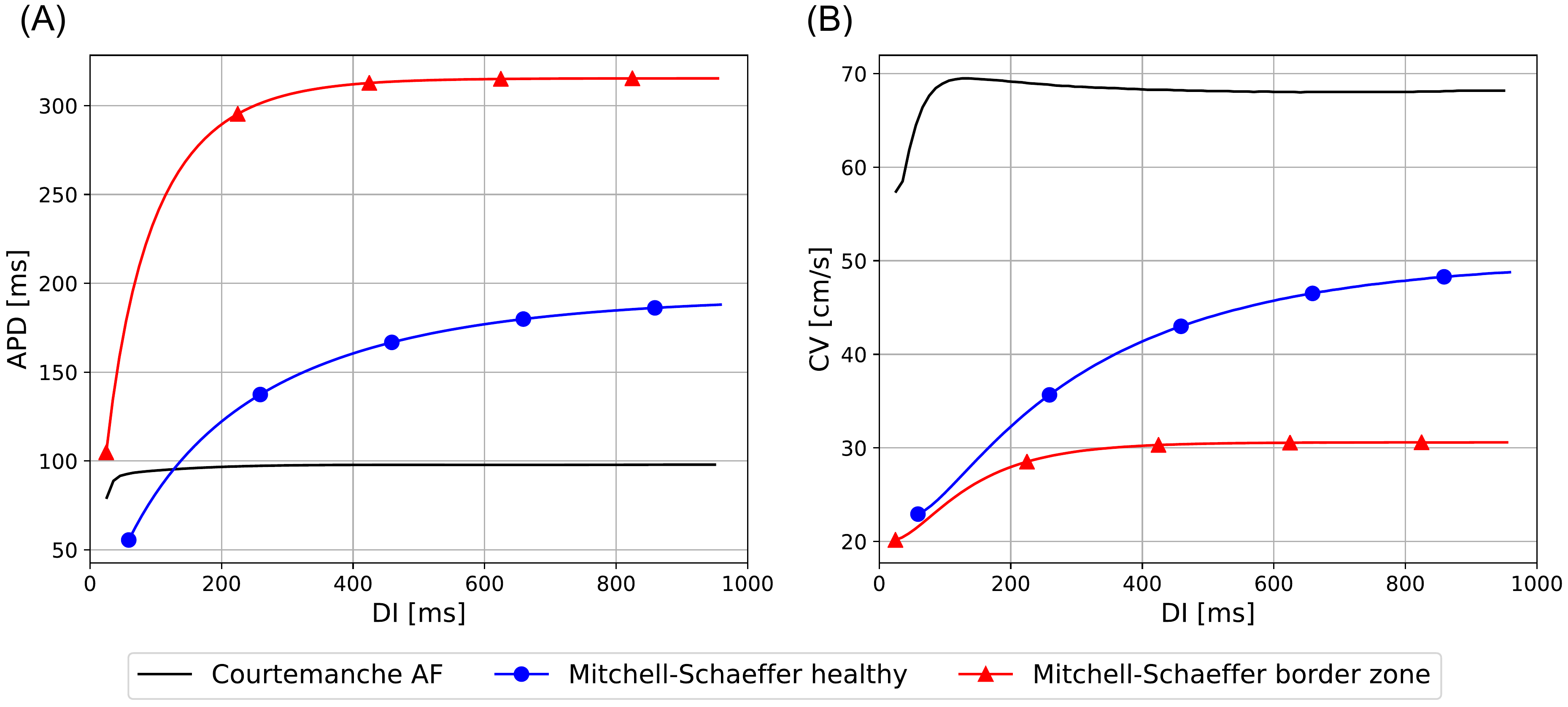}
\caption{Restitution curves of the APD (panel A) and of the CV (panel B) for the different ionic models considered.}
\label{fig: restitution}
\end{figure}

In the computation of the Courtemanche restitution curves, the CL ranges between $\text{BCL}=\SI{1}{\second}$ and $\text{CL}_0=\SI{0.1}{\second}$, and reduces with decrements of \SI{10}{\ms}. The number of prepacing stimuli is 2000, whereas 20 stimuli are delivered at BCL before measuring the restitution properties at each CL.
Moreover, the monodomain conductivity is $\sigma_\mathrm{m}=\SI{1.5}{\milli\siemens\per\cm}$~\cite{gharaviri2020epicardial}.
In the computation of the Mitchell-Schaeffer restitution curves, the prepacing is not needed, the CL ranges between $\text{BCL}=\SI{1.5}{\second}$ and $\text{CL}_0=\SI{0.2}{\second}$, reducing with decrements of \SI{10}{\ms}, and the number of stimuli delivered at BCL before measuring the restitution properties at each CL is 5.
The monodomain conductivities are set to achieve a CV of \SI{50}{\cm\per\second} in the healthy tissue and a CV of \SI{30}{\cm\per\second} in the border zone~\cite{pernod2011multi} and are respectively $\sigma_\mathrm{m}=\SI{0.46}{\milli\siemens\per\cm}$ and $\sigma_\mathrm{m}=\SI{0.19}{\milli\siemens\per\cm}$.
The restitution curves are shown in Figure~\ref{fig: restitution}.

The minimum DI for which the stimulation generates an action potential is $\text{DI}_\text{min}=\SI{35}{\ms}$ for the Courtemanche model, $\text{DI}_\text{min}=\SI{60}{\ms}$ for the Mitchell-Schaeffer model for the healthy tissue and $\text{DI}_\text{min}=\SI{25}{\ms}$ for the Mitchell-Schaeffer model for the border zone. 

\subsection{Eikonal approximation}

The eikonal approximation is given by the eikonal equation for the activation time $\phi\colon\Omega\rightarrow\mathbb{R}$, that reads \cite{colli2014mathematical}
\begin{equation}
\begin{cases}
\sqrt{\nabla\phi^\intercal\boldsymbol{D}\nabla\phi}=1,
&\text{in}\ \Omega\backslash\Omega_0,\\
\phi(\cdot)=\phi_0(\cdot),
&\text{in}\ \Omega_0,
\end{cases}
\label{eq: eikonal}
\end{equation}
where $\boldsymbol{D}\colon\Omega\rightarrow\mathbb{R}^{3\times 3}$ is the anisotropy tensor given by
\begin{equation*}
\boldsymbol{D}(\boldsymbol{x})=CV_t^2(\boldsymbol{x})\boldsymbol{I}+\big(CV_l^2(\boldsymbol{x})-CV_t^2(\boldsymbol{x})\big)\boldsymbol{f}_l(\boldsymbol{x})\boldsymbol{f}_l(\boldsymbol{x})^\intercal,
\end{equation*}
where $CV_{l,t}\colon\Omega\rightarrow\mathbb{R}$ are respectively the longitudinal and transversal conduction velocities.
The domain $\Omega_0\subset\Omega$ is the region where the action potential is initiated and $\phi_0\colon\Omega_0\rightarrow\mathbb{R}$ is the initial value.
Under suitable assumptions on the Dirichlet data, the eikonal equation has a unique viscosity solution, given by the Hopf-Lax formula \cite{lions1982generalized, bornemann2006finite, mirebeau2014anisotropic}
\[
\phi(\boldsymbol{x})=\inf_{\substack{\boldsymbol{\psi}\in C^1([0,1],\Omega),\\ \boldsymbol{\psi}(0)\in\Omega_0,\ \boldsymbol{\psi}(1)=\boldsymbol{x}}}\text{length}(\boldsymbol{\psi}),
\]
where the length of the path $\boldsymbol{\psi}$ connecting $\Omega_0$ to $\boldsymbol{x}\in\Omega\backslash\Omega_0$ is
\[
\text{length}(\boldsymbol{\psi})=\int_0^1\sqrt{\boldsymbol{\psi}^\prime(s)^\intercal\boldsymbol{D}^{-1}\big(\boldsymbol{\psi}(s)\big)\boldsymbol{\psi}^\prime(s)}\,\mathrm{d}s.
\]

\subsection{Fast marching method}

The eikonal equation~\eqref{eq: eikonal} is typically solved by algorithms that iteratively pass through the nodes of a mesh until acceptance of the activation times.
Our goal is to adapt an existing algorithm to account for the re-excitability.
A suitable algorithm has to be single-pass and has to guarantee the monotone acceptance of the nodes with respect to their activation times.
The FMM \cite{kimmel1998computing,sethian2000fast} is our choice, since it satisfies these requirements, contrary to the fast iterative method (FIM) \cite{fu2011fast} and the method presented in \cite{bornemann2006finite} based on an adaptive Gauss-Seidel iteration.
The FMM is a Dijkstra-like method that considers updates coming from within the elements of the mesh, contrary to the standard Dijkstra's method \cite{dijkstra1959note} that only considers updates from the edges.

The pseudo-code of the FMM is reported in Algorithm~\ref{alg: FMM}.
The set of nodes is denoted by $\boldsymbol{X}$ and the set of nodes in $\Omega_0$ is denoted by $\boldsymbol{X}_0\subset\boldsymbol{X}$. Moreover, for a node $\boldsymbol{x}$, $N(\boldsymbol{x})$ denotes the set of neighbors of $\boldsymbol{x}$ and $\mathcal{T}(\boldsymbol{x})$ denotes the set of elements that have $\boldsymbol{x}$ as vertex.
The anisotropy tensor $\boldsymbol{D}$ is assigned to the elements, whereas the nodes are assigned to the activation time $\phi$ and to a state (\textit{Far}, \textit{Considered} or \textit{Accepted}).
The \textit{Far} nodes have no activation time, the \textit{Considered} nodes have a temporary activation time $\tilde{\phi}$ that has not yet been accepted and the \textit{Accepted} nodes have a known activation time.
At each iteration, the \textit{Considered} node with minimum temporary activation time is marked as \textit{Accepted}, its non-\textit{Accepted} neighbors are marked as \textit{Considered} and their temporary activation time is updated.
\begin{algorithm*}
\caption{Fast marching method}
\label{alg: FMM}
\textcolor{gray}{\# Initialization}\\
Initialize $\boldsymbol{x}\in\boldsymbol{X}\backslash\boldsymbol{X}_0$ as \textit{Far} with $\phi(\boldsymbol{x})=\infty$\\
Initialize $\boldsymbol{x}_0\in\boldsymbol{X}_0$ as \textit{Accepted} with $\phi(\boldsymbol{x}_0)=\phi_0(\boldsymbol{x}_0)$\\
\vspace{0.25cm}
\textcolor{gray}{\# Update the neighbors}\\
Tag $\boldsymbol{x}_\mathrm{n}\in N(\boldsymbol{X}_0)\backslash\textit{Accepted}$ as \textit{Considered}\\
Compute $\tilde{\phi}(\boldsymbol{x}_\mathrm{n})=\mathrm{HL}\Big(\phi\big(N(\boldsymbol{x}_{\mathrm{n}})\big),\boldsymbol{D}\big(\mathcal{T}(\boldsymbol{x}_\mathrm{n})\big)\Big)$\\
\vspace{0.25cm}
\textcolor{gray}{\# Iterate over the nodes}\\
\While {$\textit{Considered}\neq\emptyset$}{
Tag $\boldsymbol{x}_\mathrm{a}=\text{argmin}_{\boldsymbol{x}\in\textit{Considered}}{\tilde{\phi}(\boldsymbol{x})}$ as \textit{Accepted} with $\phi(\boldsymbol{x}_\mathrm{a})=\tilde{\phi}(\boldsymbol{x}_\mathrm{a})$\\
Tag $\boldsymbol{x}_\mathrm{n}\in N(\boldsymbol{x}_\mathrm{a})\backslash\textit{Accepted}$ as \textit{Considered}\\
Compute $\tilde{\phi}(\boldsymbol{x}_\mathrm{n})=\mathrm{HL}\Big(\phi\big(N(\boldsymbol{x}_{\mathrm{n}})\big),\boldsymbol{D}\big(\mathcal{T}(\boldsymbol{x}_\mathrm{n})\big)\Big)$\\
}
\end{algorithm*}
The temporary activation time is computed from a local discretization of the Hopf-Lax formula \cite{mirebeau2014anisotropic}
\begin{equation}
\label{eq:hopflax_discrete}
\tilde{\phi}(\boldsymbol{x})=\min_{\boldsymbol{y}\in\partial M(\boldsymbol{x})}\left\{I_{M(\boldsymbol{x})}\phi(\boldsymbol{y})+\sqrt{(\boldsymbol{x}-\boldsymbol{y})^\intercal\boldsymbol{D}^{-1}\big(\mathcal{T}({\boldsymbol{x},\boldsymbol{y}})\big)(\boldsymbol{x}-\boldsymbol{y})}\right\},
\end{equation}
where $M(\boldsymbol{x})$ is the mesh of a small neighborhood of $\boldsymbol{x}\in\boldsymbol{X}\backslash\boldsymbol{X}_0$, $I_{M(\boldsymbol{x})}$ is the piecewise linear interpolation on $M(\boldsymbol{x})$ and $\mathcal{T}(\boldsymbol{x},\boldsymbol{y})$ denotes the element having $\boldsymbol{x}$ as vertex and $\boldsymbol{y}\in\partial M(\boldsymbol{x})$ on its boundary. The actual solution of the local minimization problem in Eq.~\eqref{eq:hopflax_discrete} is carried out analytically~\cite{GranditsGEASI2021,GeodesicBP2024}.

The convergence of the FMM to the viscosity solution of the eikonal equation~\eqref{eq: eikonal} is guaranteed if all the elements $\mathcal{T}$ of the mesh satisfy
\begin{equation}
\boldsymbol{e}_i^\intercal\boldsymbol{D}^{-1}(\mathcal{T})\boldsymbol{e}_j>0
\label{eq: acuteness}
\end{equation}
for all the edges $\boldsymbol{e}_{i,j}$ of a common face of $\mathcal{T}$.
The inequality~\eqref{eq: acuteness} is an acuteness condition with respect to the metric defined by $\boldsymbol{D}^{-1}$.
Our approach to deal with this condition is to use the anisotropic mesh adaptivity~\cite{dobrzynski2008anisotropic} to generate a mesh maximizing the number of elements that satisfy the acuteness condition with respect to the specific metrics of our applications.

\begin{figure}[t]
\centering
\includegraphics[width=\textwidth]{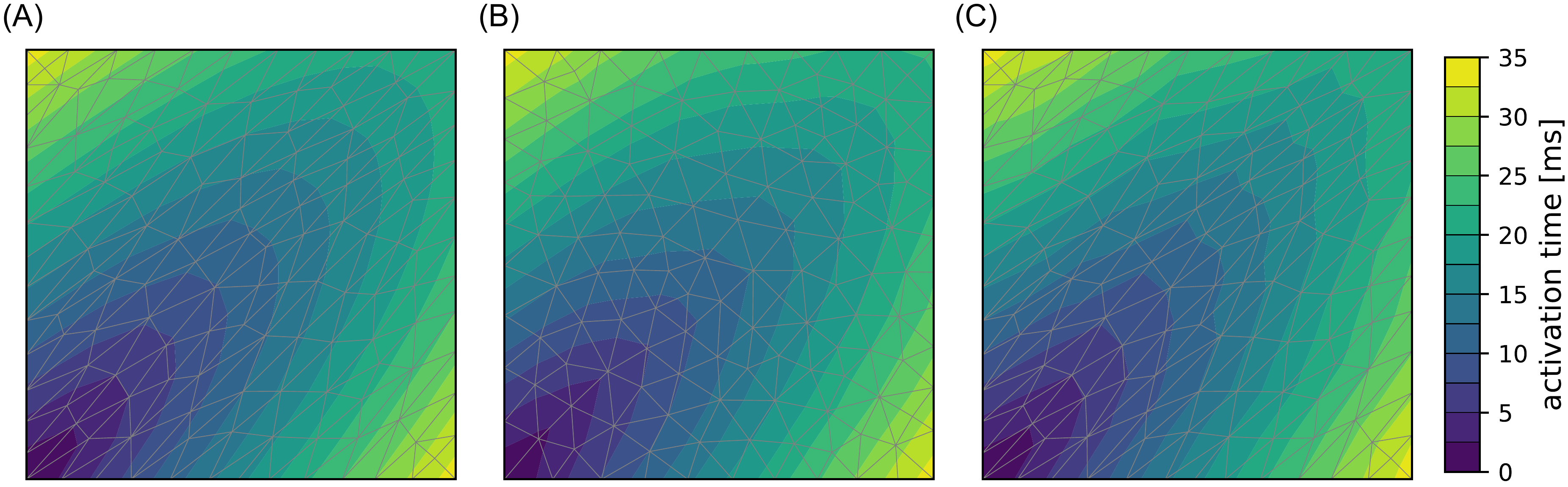}
\caption{Contour plots of the activation times obtained with the FMM on an adapted mesh (panel A), the FMM on a non-adapted mesh (panel B) and the Dijkstra's method on the adapted mesh (panel C).}
\label{fig: mesh adaptation}
\end{figure}

To illustrate the effect of the mesh adaptation and of the updates coming from within the triangles, we consider the simple case of a propagation on a square with side length of \SI{1}{\cm}. The action potential is initiated on the lower left corner, the fibers are diagonally oriented, the longitudinal CV is $CV_l=\SI{65}{\cm\per\second}$ and the anisotropy ratio is $CV_t/CV_l=0.33$. We compute the eikonal solution with the FMM on an adapted mesh, the FMM on a non-adapted mesh and the edge Dijkstra's method on the adapted mesh. The corresponding activation times are shown in Figure~\ref{fig: mesh adaptation}.

The solution obtained with the FMM on the adapted mesh is more accurate than the one computed on the non-adapted mesh and the one obtained with the Dijkstra's method. 

\subsection{Fast marching with re-excitability}

Our adaptation of the FMM to include the re-excitability requires several modifications to Algorithm~\ref{alg: FMM}.
The re-excitability is determined by the APD and the DI, which are assigned to the nodes. When a node activates, the APD determines the time interval during which it is active and the DI determines when it becomes re-excitable.
The APD depends on the DI through the restitution curves shown in Figure~\ref{fig: restitution}, panel A, and denoted by the function $a$, that depends on the tissue type of the node, which we denote by $TT$.
The $\text{DI}_\text{min}$ determining the re-excitability also depends on the tissue type.
The activation time $\phi$ is computed at each beat. In particular, it takes a finite value when the nodes are active and the value $\infty$ when the nodes are excitable.
The state \textit{Far} is assigned to the excitable nodes, the state \textit{Considered} is assigned to the excitable nodes that have non-accepted a temporary activation time $\tilde{\phi}$ and the \textit{Accepted} state is assigned to the non-excitable nodes.
To facilitate the comparison between the eikonal and the monodomain models, the nodes are also assigned to a pseudo-potential $v$ taking the value 1 when active and the value 0 when inactive.
The anisotropy tensor $\boldsymbol{D}$ is again assigned to the elements.
The longitudinal conduction velocity $CV_l$ depends on the DI through the restitution curves shown in Figure~\ref{fig: restitution}, panel B, and denoted by the function $c$, that depends on the tissue type.

The temporary activation time of a \textit{Considered} node $\boldsymbol{x}$ is computed when a neighbor $\boldsymbol{x}^\prime$ activates. When computing the temporary activation time of $\boldsymbol{x}$, the $CV_l$ of the elements having $\boldsymbol{x}$ as vertex need to be known. Since the $CV_l$ assigned to the elements depends on the DI assigned to the nodes,
a single element might take different $CV_l$ values depending on which node is being considered. Therefore we use the notation $\mathcal{T}(\boldsymbol{x},\boldsymbol{x}^\prime)$ to denote the set of elements having both $\boldsymbol{x}$ and $\boldsymbol{x}^\prime$ as vertices and to refer to the propagation from $\boldsymbol{x}^\prime$ to $\boldsymbol{x}$. The $CV_l$ assigned to the elements in $\mathcal{T}(\boldsymbol{x},\boldsymbol{x}^\prime)$ depends on the DI of $\boldsymbol{x}$ at the moment when $\boldsymbol{x}^\prime$ activates. Moreover, we use the notation $\mathcal{T}(\boldsymbol{x},:)$ to denote the set of elements having $\boldsymbol{x}$ as vertex and to refer to the propagation to $\boldsymbol{x}$.

The adapted FMM needs a global time variable in order to determine when the action potential on an activated node is over, to compute the DI and to determine when a previously activated node becomes re-excitable. These updates are not performed at each fast marching iteration, instead a time-stepping is used, i.e.~the updates are performed at the end of each time step.
The pseudo-code of the FMM with re-excitability is reported in Algorithm~\ref{alg: FMM re-excitability}.
The current time of the simulation is denoted by $t$, the time step is denoted by $\Delta t$ and the final time of the simulation is denoted by $T$.
The algorithm handles several stimuli. The number of stimuli is denoted by $N_\text{stim}$. For $i=1,...,N_\text{stim}$, the $i$-th stimulus is delivered at $\Omega_i\subset\Omega$ at time $t_i$. The set of nodes in $\Omega_i$ is denoted by $\boldsymbol{X}_i\subset\boldsymbol{X}$ for $i=1,...,N_\text{stim}$.
\begin{algorithm*}
\caption{Fast marching with re-excitability}
\label{alg: FMM re-excitability}
Initialize $\boldsymbol{x}\in\boldsymbol{X}$ as \textit{Far} with $\phi(\boldsymbol{x})=\infty$, $v(\boldsymbol{x})=0$ and $DI(\boldsymbol{x})=\infty$\\
\vspace{0.25cm}
Set $t=t_1$\\
\While {$t<T$}{
\vspace{0.25cm}
\textcolor{gray}{\# Check for stimuli}\\
\If {$t_i\in[t,t+\Delta t)$ $\mathrm{for\ some}$ $1\leq i\leq N_\mathrm{stim}$}{
Tag $\boldsymbol{x}_i\in\boldsymbol{X}_i\backslash\textit{Accepted}$ as \textit{Accepted} with $\phi(\boldsymbol{x}_i)=t_i$, $v(\boldsymbol{x}_i)=1$, $APD(\boldsymbol{x}_i)=a\big(TT(\boldsymbol{x}_i),DI(\boldsymbol{x}_i)\big)$ and $DI(\boldsymbol{x}_i)=0$\\
Tag $\boldsymbol{x}_\mathrm{n}\in N(\boldsymbol{X}_i)\backslash\textit{Accepted}$ as \textit{Considered}\\
Set $CV_l\big(\mathcal{T}(\boldsymbol{x}_\mathrm{n},\boldsymbol{x}_i)\big)=c\big(TT(\boldsymbol{x}_\mathrm{n}),DI(\boldsymbol{x}_\mathrm{n})\big)$\\
Compute $\tilde{\phi}(\boldsymbol{x}_\mathrm{n})=\mathrm{HL}\Big(\phi\big(N(\boldsymbol{x}_\mathrm{n})\big),\boldsymbol{D}\big(\mathcal{T}(\boldsymbol{x}_\mathrm{n})\big)\Big)$\\
}
\vspace{0.25cm}
\textcolor{gray}{\# Iterate within one time step}\\
\While {$\min_{\boldsymbol{x}\in\textit{Considered}}\tilde{\phi}(\boldsymbol{x})<t+\Delta t$}{
Tag $\boldsymbol{x}_\mathrm{a}=\text{argmin}_{\boldsymbol{x}\in\textit{Considered}}\tilde{\phi}(\boldsymbol{x})$ as \textit{Accepted} with $\phi(\boldsymbol{x}_\mathrm{a})=\tilde{\phi}(\boldsymbol{x}_\mathrm{a})$, $v(\boldsymbol{x}_\mathrm{a})=1$, $APD(\boldsymbol{x}_\mathrm{a})=a\big(TT(\boldsymbol{x}_\mathrm{a}),DI(\boldsymbol{x}_\mathrm{a})\big)$, $CV_l\big(\mathcal{T}(\boldsymbol{x}_\mathrm{a},:)\big)=0$ and $DI(\boldsymbol{x}_\mathrm{a})=0$\\
Tag $\boldsymbol{x}_\mathrm{n}\in N(\boldsymbol{x}_\mathrm{a})\backslash\textit{Accepted}$ as \textit{Considered}\\
Set $CV_l\big(\mathcal{T}(\boldsymbol{x}_\mathrm{n},\boldsymbol{x}_\mathrm{a})\big)=c\big(TT(\boldsymbol{x}_\mathrm{n}),DI(\boldsymbol{x}_\mathrm{n})\big)$\\
Compute $\tilde{\phi}(\boldsymbol{x}_\mathrm{n})=\mathrm{HL}\Big(\phi\big(N(\boldsymbol{x}_\mathrm{n})\big),\boldsymbol{D}\big(\mathcal{T}(\boldsymbol{x}_\mathrm{n})\big)\Big)$\\
}
\vspace{0.25cm}
\textcolor{gray}{\# Updates for the next time step}\\
Set $t=t+\Delta t$\\
For $\boldsymbol{x}\in\textit{Accepted}$ such that $\phi(\boldsymbol{x})+APD(\boldsymbol{x})<t$, set $v(\boldsymbol{x})=0$ and $DI(\boldsymbol{x})=t-\big(\phi(\boldsymbol{x})+APD(\boldsymbol{x})\big)$\\
For $\boldsymbol{x}\notin\textit{Accepted}$, set $DI(\boldsymbol{x})=DI(\boldsymbol{x})+\Delta t$\\
Tag $\boldsymbol{x}\in\textit{Accepted}$ such that $DI(\boldsymbol{x})\geq DI_\text{min}\big(TT(\boldsymbol{x})\big)$ as \textit{Far} with $\phi(\boldsymbol{x})=\infty$
}
\end{algorithm*}

\section{Numerical experiments}
\label{sec: experiments}

We test the performance of the eikonal model with re-excitability in several numerical experiments. We compare the eikonal results to the monodomain results in some simulations of re-entrant electrical activity.

The monodomain implementation is done in Python. We use the linear finite elements for the spatial discretization of the diffusion term of the monodomain equation and the implicit Euler method for the time integration of the transmembrane potential. The implicit time consists in solving a parabolic equation, with full mass matrix and interpolated currents~\cite{pezzuto2016space}. The corresponding linear system is solved in PETSc~\cite{petsc-web-page} with conjugate gradient and ILU preconditioner.
The Courtemanche ionic model is implemented in the CUDA-based Propag-5 interface~\cite{potse2006comparison,krause2012hybrid}, where, for the time integration, the Rush-Larsen method~\cite{rush1978practical} is used for the gating variables and the explicit Euler method is used for the concentration variables. Overall, the discretization is first-order accurate in time and second-order in space. The Mitchell-Schaeffer ionic model is implemented in Python, using the explicit Euler method for the time integration of the gating variable.

The FMM with re-excitability of Algorithm~\ref{alg: FMM re-excitability} is implemented in Python.
The FMM implementation relies on a min-heap structure to efficiently find the \textit{Considered} node with minimal temporary activation time at each iteration. We use the \texttt{heap} class in the \texttt{scikit-fmm} module, which allows the update of the values in the heap.

In all the numerical experiments the computational domain is discretized by a triangulated mesh with average edge length of $h=\SI{0.05}{\cm}$. The time step is $\Delta t=\SI{0.02}{\ms}$ in the monodomain simulations and $\Delta t=\SI{1}{\ms}$ in the eikonal simulations.

\subsection{Re-entries in presence of scars}
\label{sub: scars}

We first test our model in the presence of re-entrant activity around non-conductive scars, which represent the dynamics underlying VT. Therefore we consider computational domains with scars, border zone and healthy tissue. Compared to the healthy tissue, the border zone is characterized by a longer APD and a lower CV. In the monodomain model, the ionic properties are described by two Mitchell-Schaeffer ionic models. In the eikonal model, the restitution properties are described by the corresponding restitution curves, see Subsection~\ref{sub: restitution}. The longitudinal monodomain conductivities are $\sigma_\mathrm{m,h}^l=\SI{0.46}{\milli\siemens\per\cm}$ in the healthy tissue and $\sigma_\mathrm{m,bz}^l=\SI{0.19}{\milli\siemens\per\cm}$ in the border zone. The ratio between the transversal and the longitudinal monodomain conductivities is $\sigma_\mathrm{m}^t/\sigma_\mathrm{m}^l=1/9$ in both the healthy tissue and the border zone \cite{pernod2011multi}.

\subsubsection{The 2D case}
\label{par: scars 2D}

\begin{figure}[tb]
\centering
\includegraphics[width=0.5\textwidth]{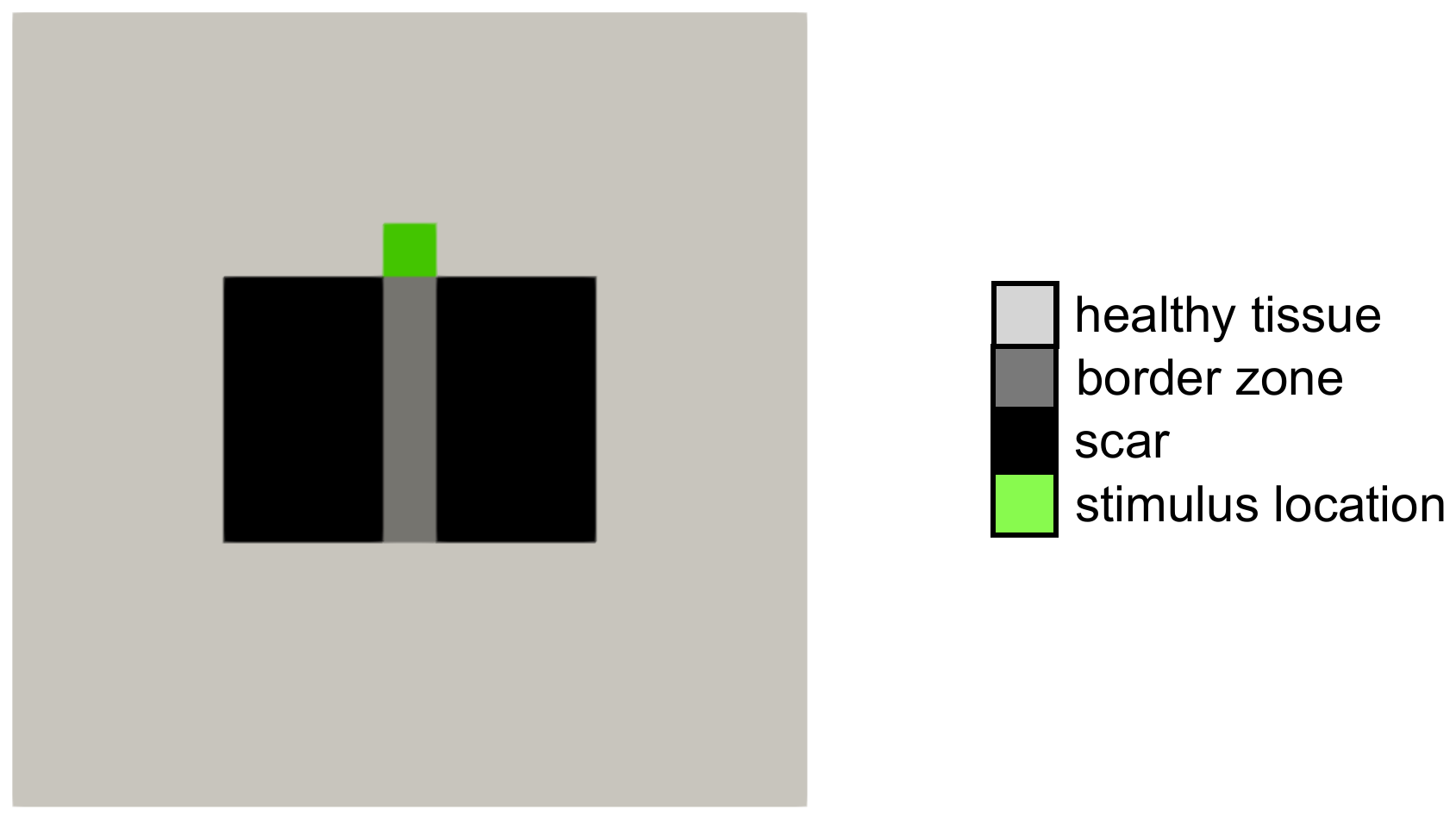}
\caption{Square with two scars, border zone and location of the stimuli.}
\label{fig: square}
\end{figure}

We first consider the 2D case of a square with side length of $\SI{15}{\cm}$. The domain has two rectangular scars of size $\SI{3}{\cm}\times\SI{5}{\cm}$ located in the middle, with a border zone of width $\SI{1}{\cm}$ between them, as illustrated in Figure~\ref{fig: square}.

We generate a mesh with the Mmg software \cite{dobrzynski2008anisotropic} and we set horizontally oriented fibers, i.e.~$\boldsymbol{f}_l=[1,0]^\intercal$.
To compute the anisotropy ratios $CV_t/CV_l$, we simulate a monodomain propagation on the healthy and the border zone tissues and we compute the CVs in the directions longitudinal and transversal to the fibers. In the healthy tissue, the longitudinal CV is $\SI{49}{\cm\per\s}$ and the transversal CV is $\SI{30}{\cm\per\s}$, thus the anisotropy ratio is $CV_t/CV_l=0.61$. In the border zone, the longitudinal CV is $\SI{30}{\cm\per\s}$ and the transversal CV is $\SI{19}{\cm\per\s}$, hence the anisotropy ratio is $CV_t/CV_l=0.63$.

The isotropic mesh that we use in the monodomain simulation is not appropriate for the eikonal simulation, as the anisotropy tensor $\boldsymbol{D}$ is not proportional to the identity $\boldsymbol{I}$. Indeed, the percentage of triangles not satisfying the acuteness condition~\eqref{eq: acuteness} with respect to the metric defined by $\boldsymbol{D}^{-1}$ is $\SI{52.7}{\percent}$. Therefore we use the Mmg software to generate an anisotropic mesh that we use in the eikonal simulation. The resulting mesh has only $\SI{0.4}{\percent}$ of triangles that do not satisfy the acuteness condition. Notice that the anisotropic mesh is comparable to the isotropic mesh in terms of number of nodes.

\begin{figure}[t]
\centering
\includegraphics[width=\textwidth]{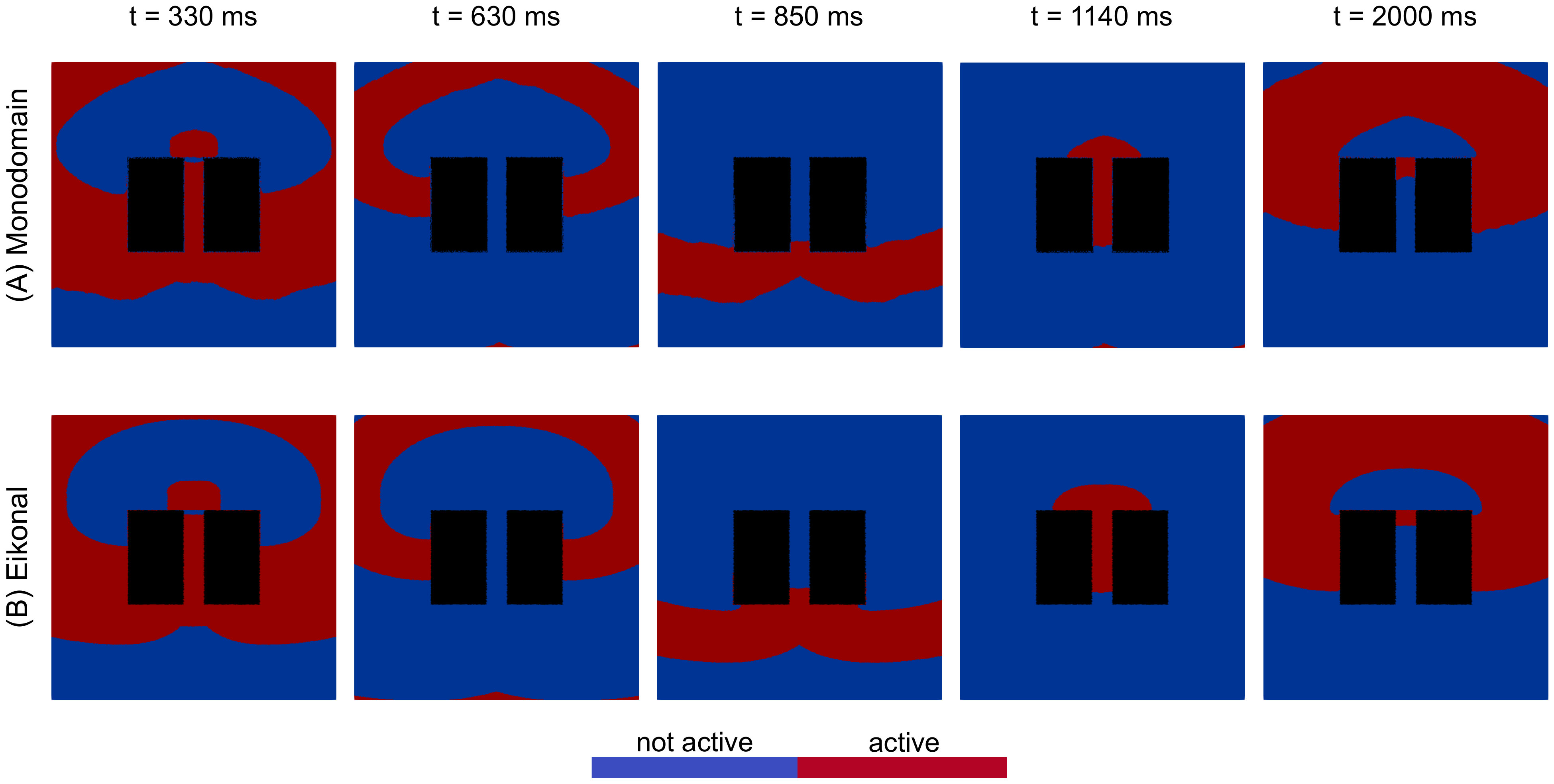}
\caption{Monodomain (panel A) and eikonal (panel B) simulations of a re-entry in presence of scars.}
\label{fig: scar}
\end{figure}

The re-entry is initiated by two stimuli of size $\SI{1}{\cm} \times \SI{1}{\cm}$ applied above the border zone, as shown in Figure~\ref{fig: square}. The first stimulus is applied at time $t=\SI{0}{\ms}$ and the second stimulus is applied at time $t=\SI{290}{\ms}$.
We perform the monodomain and the eikonal simulations on a laptop with $T=\SI{2000}{\ms}$.
The monodomain and the eikonal simulations are shown in Figure~\ref{fig: scar}.
Five snapshots show the active tissue in red and the inactive tissue in blue. In the monodomain simulation, the tissue is active when the transmembrane potential is above the $\SI{-62}{\milli\volt}$ threshold. In the eikonal simulation, the tissue is active when the pseudo-potential equals 1.
When the second stimulus is applied, the border zone is still active, as its APD is longer than the APD of the healthy tissue. As a consequence, the action potential resulting from the second stimulus can initially not propagate downwards. This dynamics is called unidirectional block. The time $t=\SI{330}{\ms}$ is shortly after the second stimulus is applied. The action potential initially propagates only upwards and laterally, and successively rotates around the scars generating a re-entry. The snapshots at the times $t=\SI{630}{\ms}$, $t=\SI{850}{\ms}$ and $t=\SI{1140}{\ms}$ illustrate the rotation and the re-entry that self-sustains until the end of the simulation at time $t=\SI{2000}{\ms}$.

The action potential propagates as a convex front. The wavefront curvature affects the CV in the monodomain model. In particular, compared to a planar front, a convex front propagates with a reduced CV \cite{fast1997role}. The eikonal model does not capture the curvature effects, as it does not account for the diffusion currents. Therefore, the action potential propagates slightly faster in the eikonal simulation than in the monodomain simulation.
The difference between the monodomain and the eikonal shape of the depolarization and repolarization fronts is due to the effect of the space resolution on the monodomain simulation, see Paragraph~\ref{par: discretization monodomain}.

\subsubsection{Ventricular tachycardia}
\label{par: VT}

\begin{figure}[t]
\centering
\includegraphics[width=0.5\textwidth]{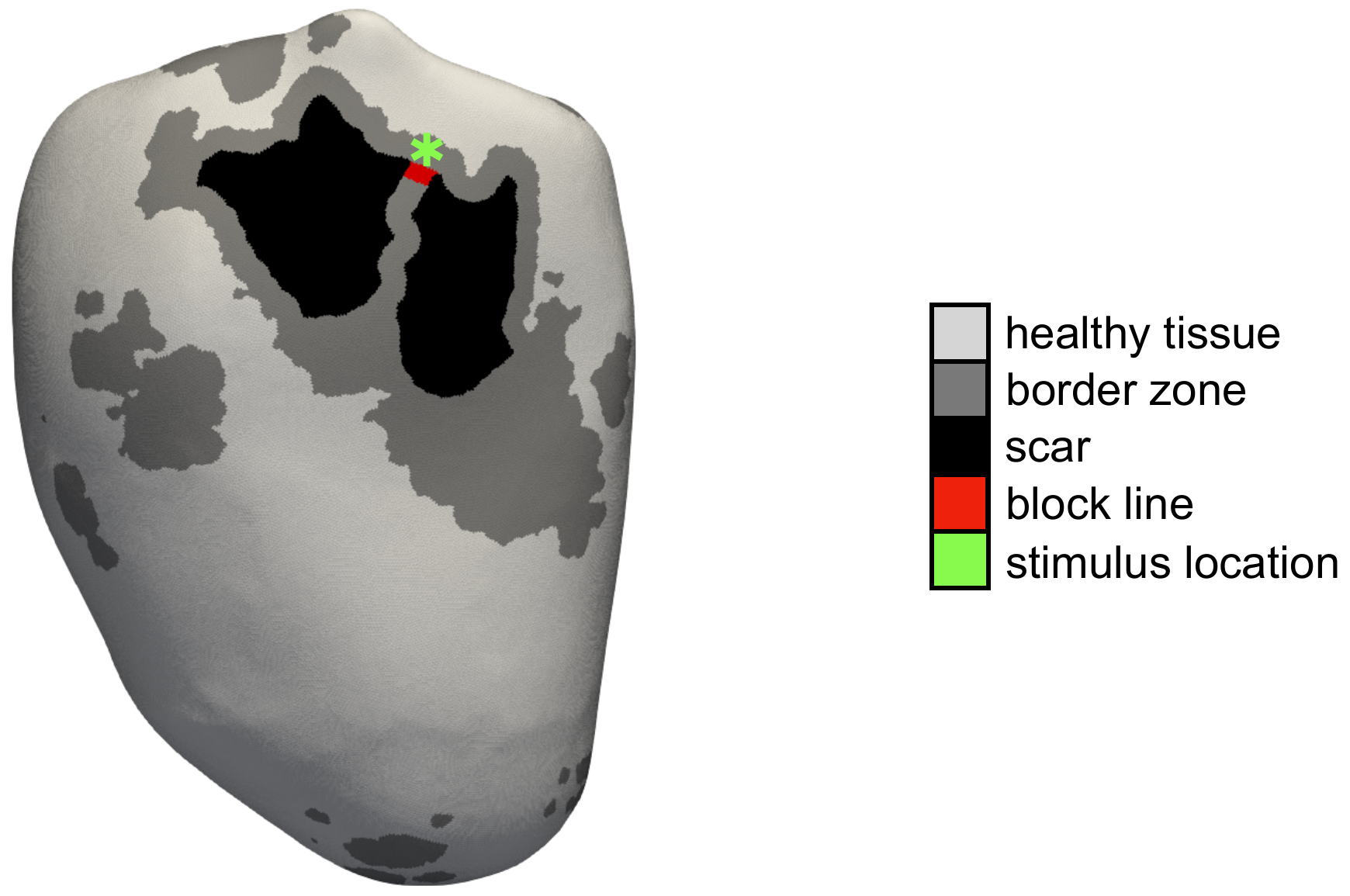}
\caption{Left ventricle with two scars, border zones, block line and location of the stimulus.}
\label{fig: ventricle}
\end{figure}

We now consider the case of the left ventricular surface with two scars and with border zones as illustrated in Figure~\ref{fig: ventricle}.
We use the Mmg software \cite{dobrzynski2008anisotropic} to refine the original coarse mesh, we map the tissue types from the coarse to the fine mesh and we set the circumferential fiber orientation as in the midwall.
To compute the anisotropy ratios $CV_t/CV_l$ for the surface, we simulate two monodomain propagations on a square, similarly to Paragraph~\ref{par: scars 2D}, but here with diagonally oriented fibers, see Paragraph~\ref{par: discretization ratio} for the motivation. In the healthy tissue, the longitudinal CV is $\SI{49}{\cm\per\s}$ and the transversal CV is $\SI{22}{\cm\per\s}$, thus the anisotropy ratio is $CV_t/CV_l=0.45$. In the border zone, the longitudinal CV is $\SI{30}{\cm\per\s}$ and the transversal CV is $\SI{14}{\cm\per\s}$, hence the anisotropy ratio is $CV_t/CV_l=0.47$.
The anisotropy tensor $\boldsymbol{D}$ is again not proportional to the identity $\boldsymbol{I}$. Therefore the isotropic mesh that we use in the monodomain simulation is not suitable for the eikonal simulation, indeed $\SI{64.0}{\percent}$ of the triangles do not satisfy the acuteness condition. Instead, the anisotropic mesh generated with the Mmg software is suitable, as the percentage of triangles not satisfying the acuteness condition is only $\SI{3.4}{\percent}$.
Thus we map the tissue types from the isotropic to the anisotropic mesh and we assign the circonferential fiber orientation also to the anisotropic mesh.

\begin{figure}[t]
\centering
\includegraphics[width=\textwidth]{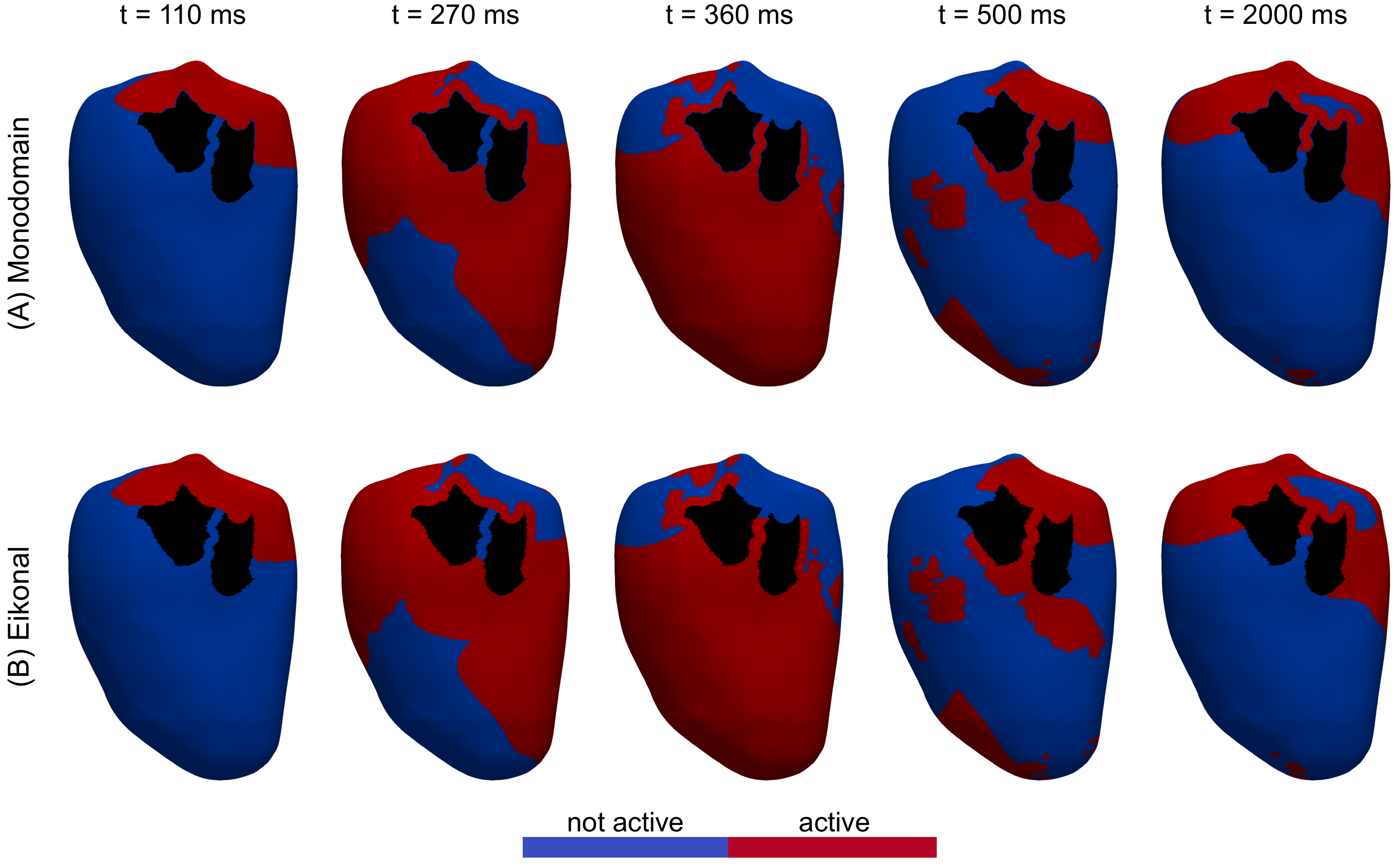}
\caption{Monodomain (panel A) and eikonal (panel B) simulations of VT.}
\label{fig: VT}
\end{figure}

We apply a single stimulation and we use the block line approach to simulate a unidirectional block that initiates VT. The block line and the location of the stimulus are shown in Figure~\ref{fig: ventricle}. The stimulus is applied at time $t=\SI{0}{\ms}$. The block line is non-conductive at the beginning of the simulation and becomes conductive at time $t=\SI{330}{\ms}$.
We perform the monodomain and the eikonal simulations on a laptop with $T=\SI{2000}{\ms}$.
The monodomain and the eikonal simulations are shown in Figure~\ref{fig: VT}.

When the stimulus is applied, because of the block, the resulting action potential cannot propagate in the gap between the two scars. As a consequence, the action potential propagates around the scars, as illustrated by the snapshots at the times $t=\SI{110}{\ms}$ and $t=\SI{270}{\ms}$. The action potential reaches the location of the block line around the time $t=\SI{360}{\ms}$, when the tissue is already conductive. Therefore, the action potential can continue the rotation around the scars, as shown in the snapshot at time $t=\SI{500}{\ms}$. This re-entrant dynamics self-sustains until the end of the simulation at time $t=\SI{2000}{\ms}$.
Again, the eikonal action potential propagation is slightly faster than the monodomain propagation. This is again due to the fact that the wavefront is mostly convex and that the eikonal model does not capture the effect of the curvature.

\subsection{Spiral waves}
\label{sub: spiral waves}

The last numerical experiments are simulations of spiral waves, which characterize the dynamics of an atrial flutter (AFlut). Therefore we consider computational domains of healthy atrial tissue. In the monodomain model, the ionic properties are described by the Courtemanche ionic model adapted to the AF electric remodeling. In the eikonal model, the restitution properties are described by the corresponding restitution curves, see Subsection~\ref{sub: restitution}. In the monodomain simulations, the initial conditions on $v$ and $\boldsymbol{w}$ are obtained after the simulation of 2000 beats at $\text{BCL}=\SI{1}{\s}$ in a single-cell model. The longitudinal and transversal monodomain conductivities are respectively $\sigma_\mathrm{m}^l=\SI{1.5}{\milli\siemens\per\cm}$ and $\sigma_\mathrm{m}^t=\SI{0.24}{\milli\siemens\per\cm}$ \cite{gharaviri2020epicardial}.

\subsubsection{The 2D case}
\label{par: spiral 2D}

We first consider the 2D case of a square with side length of $\SI{15}{\cm}$ and horizontally oriented fibers, i.e.~$\boldsymbol{f}_l=[1,0]^\intercal$.
The isotropic mesh is the same as in Paragraph~\ref{par: scars 2D}.
We compute the anisotropy ratio $CV_t/CV_l$ with a monodomain simulation as described in Paragraph~\ref{par: scars 2D}. The longitudinal CV is $\SI{68}{\cm\per\s}$, whereas the transversal CV is $\SI{34}{\cm\per\s}$. Therefore the anisotropy ratio is $CV_t/CV_l=0.5$.
The isotropic mesh for the monodomain simulation has $\SI{74.8}{\percent}$ of triangles that do not satisfy the acuteness condition with respect to the metric defined by $\boldsymbol{D}^{-1}$, therefore it is not appropriate for the eikonal simulation. Instead the anisotropic mesh generated with the Mmg software \cite{dobrzynski2008anisotropic} is appropriate, as the percentage of triangles not satisfying the acuteness condition is only $\SI{0.4}{\percent}$.

Similarly to \cite{gassa2021spiral}, the spiral wave is initiated by a first stimulus with width $\SI{0.1}{\cm}$ on the left side of the domain at time $t=\SI{0}{\ms}$, followed by a second stimulus on the lower left rectangle of size $\SI{6}{\cm}\times \SI{12}{\cm}$ at time $t=\SI{210}{\ms}$.
We perform the monodomain and the eikonal simulations on a laptop with $T=\SI{1000}{\ms}$.

\begin{figure}[t]
\centering
\includegraphics[width=\textwidth]{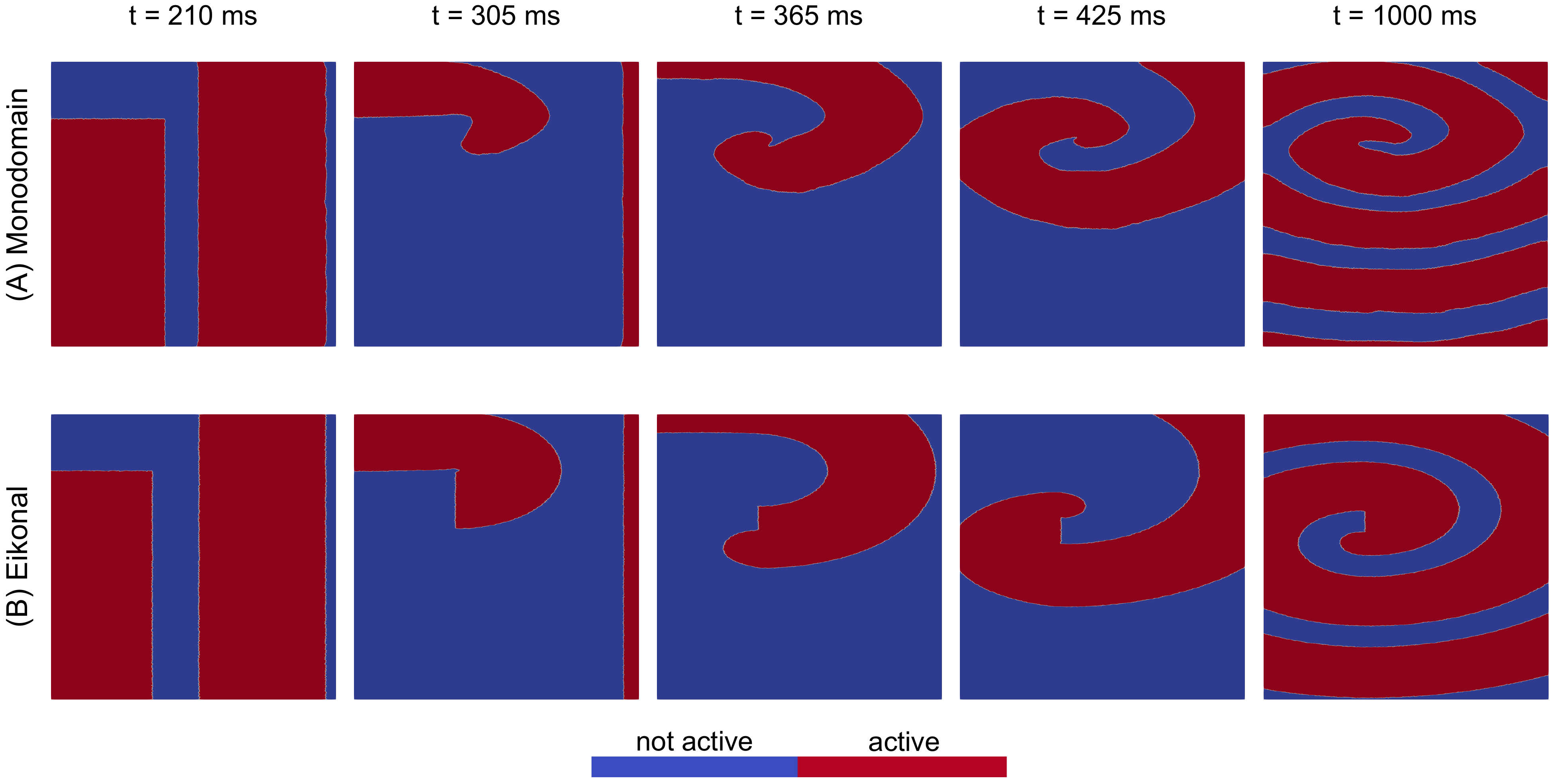}
\caption{Monodomain (panel A) and eikonal (panel B) simulations of a spiral wave.}
\label{fig: spiral}
\end{figure}

The monodomain and the eikonal simulations are shown in Figure~\ref{fig: spiral}.
The time $t=\SI{210}{\ms}$ is the instant when the second stimulus is applied. Note that part of the stimulated tissue might not be excitable. In the monodomain case, the stimulation in the non-excitable tissue results in a quick increase of the transmembrane potential, that then rapidly decreases back to its pre-stimulation value, without affecting the overall dynamics. In the eikonal case, because of the implementation of algorithm~\ref{alg: FMM re-excitability}, the non-excitable tissue is not activated.
The action potential resulting from the second stimulus initially propagates only upwards and successively generates a re-entry.
At time $t=\SI{305}{\ms}$, the tissue where the second stimulus was applied is inactive, but it is not yet excitable, so the action potential only propagates towards the right and downwards.
At time $t=\SI{365}{\ms}$, the stimulated tissue is excitable and the action potential propagates also towards the left.
The re-entry, generated around the time $t=\SI{425}{\ms}$, induces a spiral wave that self-sustains until the end of the simulation at time $t=\SI{1000}{\ms}$.
The action potential resulting from the first stimulus propagates as a planar front, therefore the eikonal model is accurate in describing the related depolarization and repolarization.
Instead, in the spiral wave resulting from the second stimulus there are differences between the simulations that are due to the effect of the wavefront curvature on the monodomain CV, which is not captured by the eikonal model.
On the one hand, since the action potential propagates as a convex front, the eikonal propagation is faster than the monodomain propagation \cite{fast1997role}.
On the other hand, the monodomain rotation is faster than the eikonal rotation, as the tip of the re-entry moves in the monodomain simulation, whereas it is fixed in the eikonal simulation \cite{panfilov2018theory}.
Again, the shape of the depolarization and repolarization fronts is different in the monodomain and the eikonal cases because of the effect of the space resolution on the monodomain simulation, see Paragraph~\ref{par: discretization monodomain}.

\subsubsection{Atrial flutter}
\label{par: AFlut}

\begin{figure}[t]
\centering
\includegraphics[width=0.5\textwidth]{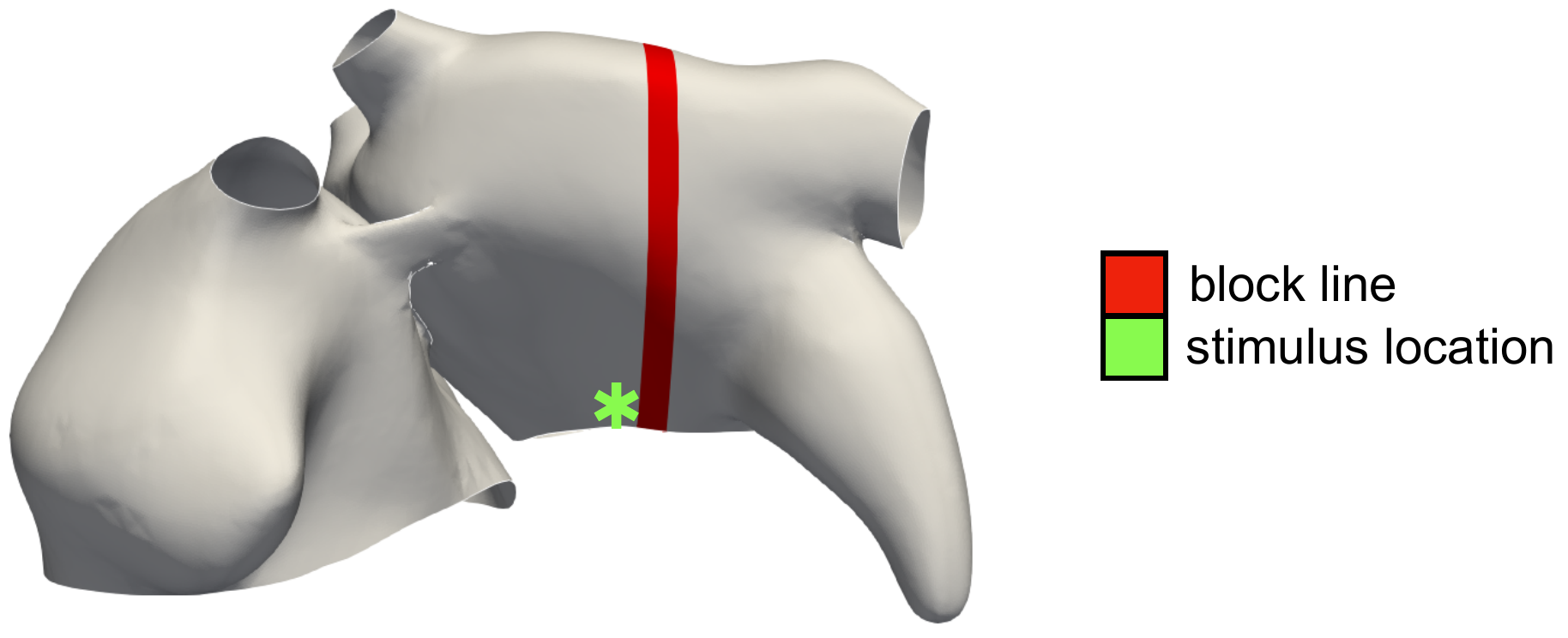}
\caption{Atria with block line and location of the stimulus.}
\label{fig: atria}
\end{figure}

We now consider the case of the atrial surface illustrated in Figure~\ref{fig: atria}.
We use the Mmg software \cite{dobrzynski2008anisotropic} to refine the original coarse mesh and we set the fiber orientation by projecting the horizontal vector $[1,0,0]^\intercal$ on the triangles.
We compute the anisotropy ratio $CV_t/CV_l$ for the surface with a monodomain simulation as described in Paragraph~\ref{par: VT} and motivated by the observations in Paragraph~\ref{par: discretization ratio}. The longitudinal CV is $\SI{68}{\cm\per\s}$, whereas the transversal CV is $\SI{27}{\cm\per\s}$. Therefore the anisotropy ratio is $CV_t/CV_l=0.4$.
This result is in agreement with the formulas \cite{colli2014mathematical}
\begin{equation*}
CV_l=\frac{\rho\sqrt{\sigma_\mathrm{m}^l}}{C_\mathrm{m}\sqrt{\beta}} \quad\text{and}\quad CV_t=\frac{\rho\sqrt{\sigma_\mathrm{m}^t}}{C_\mathrm{m}\sqrt{\beta}},
\end{equation*}
where $\rho\in\mathbb{R}$ is a parameter that depends on the ionic model. Indeed $CV_t/CV_l=\sqrt{\sigma_\mathrm{m}^t/\sigma_\mathrm{m}^l}=0.4$.
Again, the isotropic mesh for the monodomain simulation is not suitable for the eikonal simulation, as the percentage of triangles not satisfying the acuteness condition is $\SI{75.1}{\percent}$. Instead, the anisotropic mesh generated with the Mmg software has $\SI{8.8}{\percent}$ of triangles that do not satisfy the acuteness condition and is more suitable.
Hence we assign the fiber orientation also to the anisotropic mesh by projecting the horizontal vector on the triangles.

\begin{figure}[t]
\centering
\includegraphics[width=\textwidth]{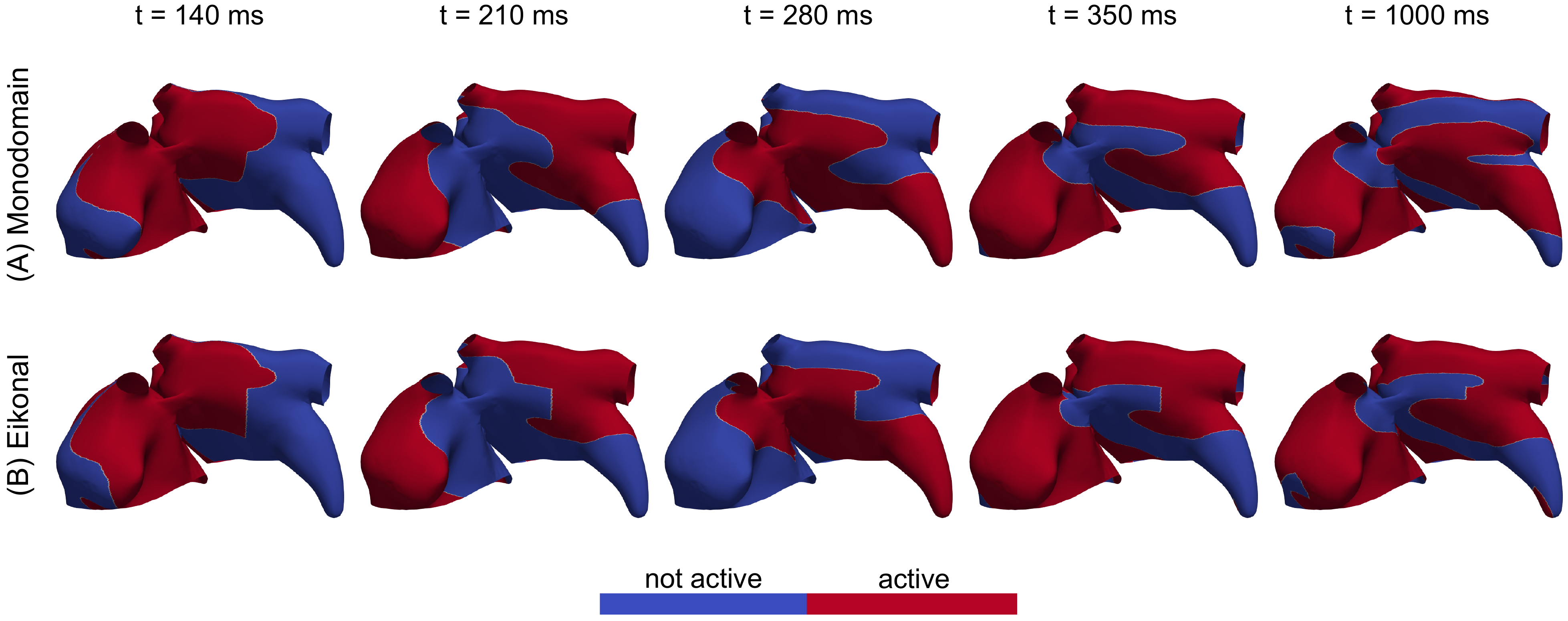}
\caption{Monodomain (panel A) and eikonal (panel B) simulations of AFlut.}
\label{fig: AFlut}
\end{figure}

AFlut is initiated by a single stimulation with the block line approach \cite{gharaviri2020epicardial}. The block line and the stimulation site are shown in Figure~\ref{fig: atria}. The stimulus is applied at time $t=\SI{0}{\ms}$. The block line is initially non-conductive and becomes conductive at time $t=\SI{125}{\ms}$.
The monodomain and the eikonal simulations are shown in Figure~\ref{fig: AFlut}.
The action potential initially propagates upwards and towards the left. The time $t=\SI{140}{\ms}$ is shortly after the block line became conductive and the action potential propagates also towards the right and downwards. Around the time $t=\SI{210}{\ms}$ the left side of the block line becomes re-excitable and the action potential propagates again towards the left. This re-entrant dynamics repeats, as illustrated by the snapshots at the times $t=\SI{280}{\ms}$ and $t=\SI{350}{\ms}$, and self-sustains until the end of the simulation at time $t=\SI{1000}{\ms}$.
Again, because the eikonal model does not capture the effect of the wavefront curvature, there are differences between the simulated re-entries.
The tip of the re-entry is fixed in the eikonal simulation, whereas it moves in the monodomain simulation. As a consequence, the monodomain rotation is faster than the eikonal rotation.
However, the eikonal propagation is faster than the monodomain propagation, since the action potential propagates as a convex front.

\section{Discussion}
\label{sec: discussion}


In this work, we presented a fast eikonal model with re-excitability for simulations of cardiac arrhythmias. Our model is based on the FMM and handles the anisotropy by mesh adaptation. We compared the eikonal model to the monodomain model in several numerical experiments.
The eikonal model is very accurate for simulations of re-entries that rotate around scars.
Our approach can also simulate focal re-entries whose spiral tip is free to move in the domain, with good accuracy compared to monodomain simulations.
The differences observed between the eikonal and monodomain models are mainly due to the effect of the wavefront curvature \cite{fast1997role} on the CV, which is not captured by the eikonal model.

The computational cost of the monodomain model depends on the size of the linear system resulting from the discretization of the monodomain equation and on the complexity of the ionic model. Instead, the computational cost of the eikonal model is proportional to the number of node updates. Therefore, the speedup of the eikonal model compared to the monodomain model varies depending on the numerical experiment.
In our numerical experiments, our implementation of the eikonal model is between 7 and 10 times faster than the implementation of the monodomain model.
Note that, in our numerical experiments, the meshes used for the monodomain and the eikonal simulations are comparable in terms of number of nodes. However, the FMM can be applied on coarser meshes, since, contrary to the numerical methods used to solve the monodomain system, it has no stability issues \cite{colli2014mathematical}. Clearly, a coarser spatial discretization in the eikonal model would lead to a higher speedup compared to the monodomain model. Furthermore, our monodomain implementation is highly optimized and ionic models are executed on GPU.

Our Python implementation of the FMM with re-excitability is not optimized. To compare our prototype to other eikonal implementations, we consider the case of a propagation with CV of $\SI{68}{\cm\per\s}$ on a squared tissue with side length of $\SI{15}{\cm}$ discretized with space resolution $h=\SI{0.05}{\cm}$. The computing time of the FMM operating on a grid implemented in the C++ based \texttt{scikit-fmm} module is $\SI{0.04}{\s}$, whereas the computing time of the FIM operating on a triangulated mesh implemented in the Python based \texttt{fim-python} module is $\SI{6.62}{\s}$. The computing time of our implementation is $\SI{42.65}{\s}$ for the standard FMM of Algorithm~\ref{alg: FMM} and $\SI{83.86}{\s}$ for the FMM with re-excitability of Algorithm~\ref{alg: FMM re-excitability}. The maximum activation time in this numerical experiment is $\SI{0.32}{\s}$, thus the FMM implementation provided by the \texttt{scikit-fmm} module is better than real-time.
An optimized Python implementation of the FMM on triangulated meshes is at least as fast as the implementation of the FIM provided by the \texttt{fim-python} module, hence the efficiency of our prototype can be improved at least by a factor $6.5$.
Moreover, the adaptations to the FMM required to include the re-excitability do not change the computational complexity, which is $O(N\log N)$ \cite{kimmel1998computing,sethian2000fast}, where $N$ is the number of nodes of the mesh.
Therefore, we believe that an optimized C++ implementation of the FMM is potentially real-time and that the computational overhead due to the modifications needed to include the re-excitability is negligible \cite{sermesant2007anisotropic}.
However, the optimization of the implementation is beyond the scope of this paper and is part of our future work.

Currently there is a compelling need of computational models based on imaging data that meet practical requirements when employed to non-invasively study cardiac arrhythmias and plan their treatment. These requirements are a computational cost within the clinical time constraints and the ability to operate without off-site high-performance computing resources \cite{campos2022automated}.
The eikonal model with re-excitability presented in this work is potentially real-time and operates on desktop computational resources.
In order to be applied to patient-specific cases, the eikonal model needs a training phase. Similarly to the monodomain model, the training requires the mapping of the tissue properties to the computational mesh. Additionally, the training of the eikonal model requires the CV tuning, the mesh adaptation and the computation of the restitution curves.
The computational cost of the training phase is negligible compared to the cost of performing several monodomain simulations. Often the \textit{in-silico} planning of the ablation treatment needs the evaluation of the arrhythmia inducibility after pacing from several locations \cite{gander2022fast,campos2022automated}. Therefore, in such cases, it is worth investing the computational effort of the training phase to exploit the real-time potential of the eikonal model in the electrophysiological simulations, instead of using the monodomain model.

In this work, our computational domains consisted of triangulated surfaces. The FMM and the modifications to include the re-excitability can straightforwardly be applied to 3D domains discretized by tetrahedra. The mesh adaptation to handle the anisotropy can be performed with the Mmg software \cite{dobrzynski2008anisotropic} also on tetrahedral elements. However, since tetrahedra have more angles than triangles, in order to obtain a mesh with a high percentage of elements that satisfy the acuteness condition, one would need to increase the spatial resolution, thus increasing the computational cost.
Another limitation of this work is the fact that we did not consider the presence of fibrosis. The discontinuities in the conductivity that model the fibrotic tissue introduce delays in the activation times. The discretized monodomain model amplifies these delays, whereas the eikonal model does not capture them \cite{gander2023accuracy}. Therefore, the presence of fibrosis affects the accuracy of both the monodomain and the eikonal models. To overcome this difficulty, one could employ the eikonal-diffusion model \cite{colli2014mathematical}, which accurately describes the delays \cite{gander2023accuracy}. Additionally, the eikonal-diffusion model captures the effects of the wavefront curvature on the CV, similarly to the eikonal-curvature model \cite{keener1991eikonal,colli2014mathematical}. However, the applicability of the eikonal-diffusion and the eikonal-curvature models to simulations of re-entries needs to be studied in the future, both in terms of algorithmical inclusion of the re-excitability and in terms of computational cost.


\bibliographystyle{elsarticle-harv}
\bibliography{biblio}

\section*{Funding}
SP and RK acknowledges the support of the SNF ``CardioTwin'' project (no.~214817) and the CSCS-Swiss National Supercomputing Centre project no.~s1275. SP also acknowledges the PRIN-PNRR project no.~P2022N5ZNP and INdAM-GNCS. FSC acknowledges the support of the project ANID - ERAPERMED-13  and ANID – Millennium Science Initiative Program – ICN2021\_004.

\section*{Author contributions}
\textbf{LG}: Conceptualization, Methodology, Software, Validation, Formal analysis, Investigation, Data curation, Writing -- original draft, Writing -- review \& editing, Visualization.
\textbf{FSC}: Methodology, Resources, Writing -- original draft, Writing -- review \& editing, Funding acquisition.
\textbf{RK}: Conceptualization, Resources, Writing -- original draft, Writing -- review \& editing, Project administration, Funding acquisition.
\textbf{SP}: Conceptualization, Methodology, Resources, Writing -- original draft, Writing -- review \& editing, Project administration, Funding acquisition.

\section*{Competing interests}
None

\appendix

\section{Discretization issues}

In Subsections~\ref{sub: scars} and~\ref{sub: spiral waves}, the monodomain CVs are affected by the spatial discretization. In this Subsection we analyze the influence of the spatial discretization on the monodomain simulations and on the anisotropy ratios.

\subsection{Monodomain simulations}
\label{par: discretization monodomain}

\begin{figure}[t]
\centering
\includegraphics[width=\textwidth]{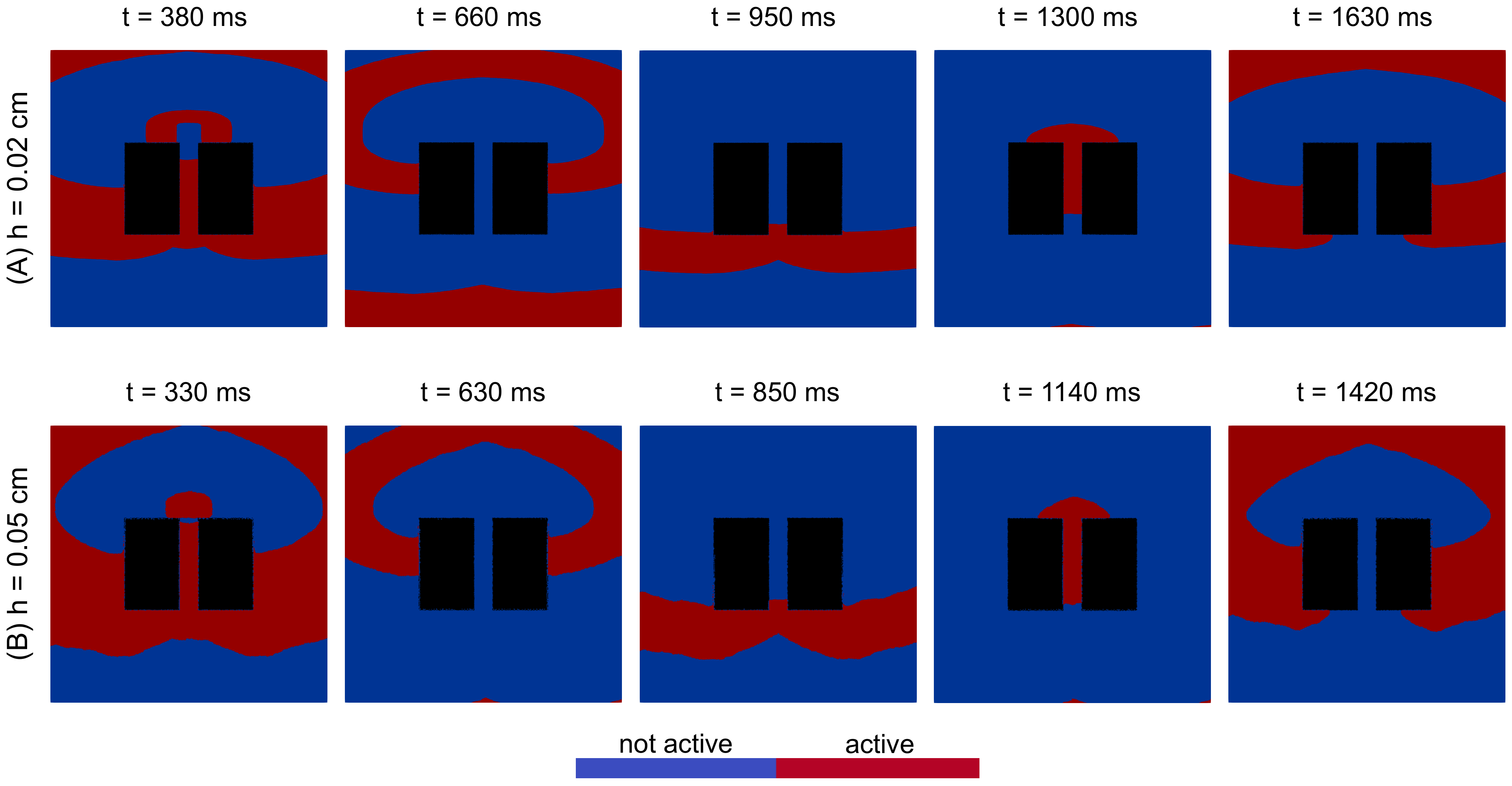}
\caption{Monodomain simulations of a re-entry in presence of scars with space resolution $h=\SI{0.02}{\cm}$ (panel A) and $h=\SI{0.05}{\cm}$ (panel B).}
\label{fig: discretization}
\end{figure}

The space resolution $h=\SI{0.05}{\cm}$ used in Subsections~\ref{sub: scars} and~\ref{sub: spiral waves} affects the monodomain simulations. As a comparison, we perform the monodomain simulation of Paragraph~\ref{par: scars 2D} with the finer resolution $h=\SI{0.02}{\cm}$. The comparison is shown in Figure~\ref{fig: discretization}.
Since we use finite elements, the finer spatial discretization has the effect of reducing the CV \cite{pezzuto2016space}. Therefore, in Figure~\ref{fig: discretization}, the two simulations are illustrated by snapshots taken at different times.
In the simulation with the finer resolution, the ellipsoidal shape of the depolarization and repolarization fronts is more pronounced than in the simulation with the coarser resolution.
In fact, when the mesh is too coarse, the front shape is affected by the mesh topology.
Contrary to the monodomain model, the eikonal model captures the ellipsoidal shape of the fronts also at resolution $h=\SI{0.05}{\cm}$, see Figure~\ref{fig: scar}.
Notice that the same observations hold also for the simulations of Paragraph~\ref{par: spiral 2D} shown in Figure~\ref{fig: spiral}.

\subsection{Anisotropy ratios}
\label{par: discretization ratio}

\begin{figure}[t]
\centering
\includegraphics[width=\textwidth]{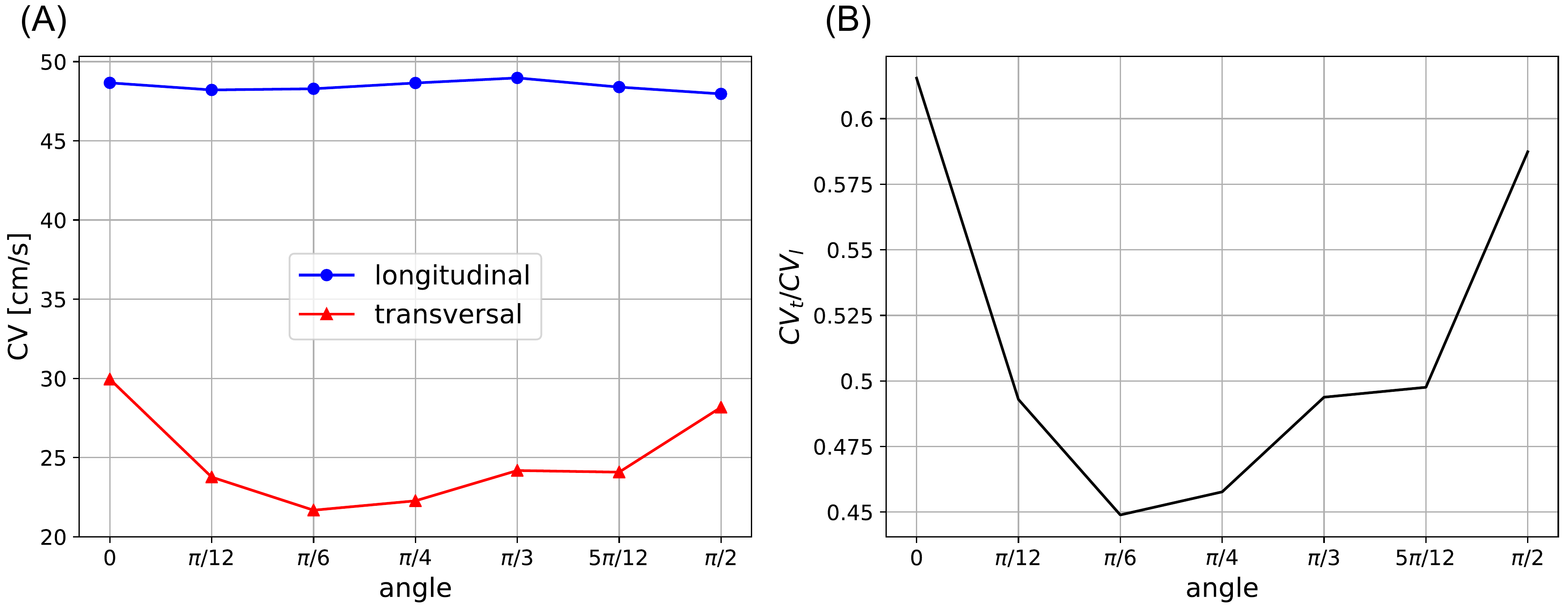}
\caption{Longitudinal and transversal CVs (panel A) and anisotropy ratio (panel B) vs.~angle of the fiber orientation with respect to the horizontal axis.}
\label{fig: anisotropy}
\end{figure}

The spatial discretization affects the monodomain CVs and, as a consequence, the anisotropy ratios. To illustrate this fact, we consider the isotropic triangulated mesh of Paragraphs~\ref{par: scars 2D} and~\ref{par: spiral 2D}, and fiber orientations with variable angles with respect to the horizontal axis. For all the fiber directions, we simulate a monodomain propagation with the Mitchell-Schaeffer ionic model for the healthy ventricular tissue and we compute the longitudinal and transversal CVs and the corresponding anisotropy ratios. The results are shown in Figure~\ref{fig: anisotropy} for seven angles varying between $0$ and $\pi/2$.
The longitudinal CV is approximately constant, whereas the transversal CV, and consequently the anisotropy ratio, is larger for horizontal and vertical fibers than for diagonal fibers.
Since the coarser spatial discretization has the effect of increasing the CV \cite{pezzuto2016space}, these results suggest that the triangles of the mesh are mostly oriented along the horizontal and vertical axes, thus implying a coarser resolution along these directions.
On a mesh discretizing a surface, it is unlikely that the fiber direction matches the orientation of the triangles. Therefore, it is suitable to set diagonal fibers when approximating the CVs on a mesh discretizing a square. This motivates our choice for the computation of the anisotropy ratios in Paragraphs~\ref{par: VT} and~\ref{par: AFlut}.

\end{document}